\documentclass[runningheads]{llncs}
\usepackage[T1]{fontenc}
\usepackage{graphicx}
\usepackage{amssymb}
\usepackage{multirow}
\usepackage{booktabs}
\usepackage{subfigure}
\usepackage{subfloat}
\usepackage{tabularx} 
\usepackage{caption}
\usepackage{amsmath}
\usepackage[misc]{ifsym}
\begin{document}

\title{TransUKAN:Computing-Efficient Hybrid KAN-Transformer for Enhanced Medical Image Segmentation}

\author{Yanlin Wu\inst{1} \and Tao Li \inst{1,2}$^{(\textrm{\Letter})}$ \and Zhihong Wang \inst{1} \and  Hong Kang \inst{1} \and Along He \inst{1}}
\authorrunning{F. Author et al.}

\institute{Tianjin
	Key Laboratory of Network and Data Security Technology, College of Computer Science, Nankai University, Tianjin, China \\ 
		 \email{litao@nankai.edu.cn}\\ \and
	Haihe Lab of ITAI}

\maketitle

\begin{abstract}
U-Net is currently the most widely used architecture for medical image segmentation. Benefiting from its unique encoder-decoder architecture and skip connections, it can effectively extract features from input images to segment target regions. The commonly used U-Net is typically based on convolutional operations or Transformers, modeling the dependencies between local or global information to accomplish medical image analysis tasks. However, convolutional layers, fully connected layers, and attention mechanisms used in this process introduce a significant number of parameters, often requiring the stacking of network layers to model complex nonlinear relationships, which can impact the training process. To address these issues, we propose TransUKAN. Specifically, we have improved the KAN to reduce memory usage and computational load. On this basis, we explored an effective combination of KAN, Transformer, and U-Net structures. This approach enhances the model's capability to capture nonlinear relationships by introducing only a small number of additional parameters and compensates for the Transformer structure's deficiency in local information extraction. We validated TransUKAN on multiple medical image segmentation tasks. Experimental results demonstrate that TransUKAN achieves excellent performance with significantly reduced parameters. 
The code will be available athttps://github.com/wuyanlin-wyl/TransUKAN.
\end{abstract}

\section{Introduction}
\label{sec:introduction}
Medical image segmentation is a crucial task in medical imaging analysis, aiming to accurately segment different anatomical structures or lesion areas in images \cite{chen2024learning,cheng2024few,shaker2024unetr++}. This process plays a vital role in clinical diagnosis, surgical planning, and treatment evaluation \cite{tian2023self,bilic2023liver,liu2024cosst}. Traditional medical image segmentation methods typically rely on manual feature extraction and heuristic rules. However, these methods exhibit significant limitations when dealing with complex and diverse medical images \cite{demirhan2014segmentation,yang2017lung}. With the development of deep learning technology, data-driven approaches have gradually become the mainstream in the field of medical image segmentation.

\begin{figure}[t]
  \centering
    \includegraphics[width=\columnwidth]{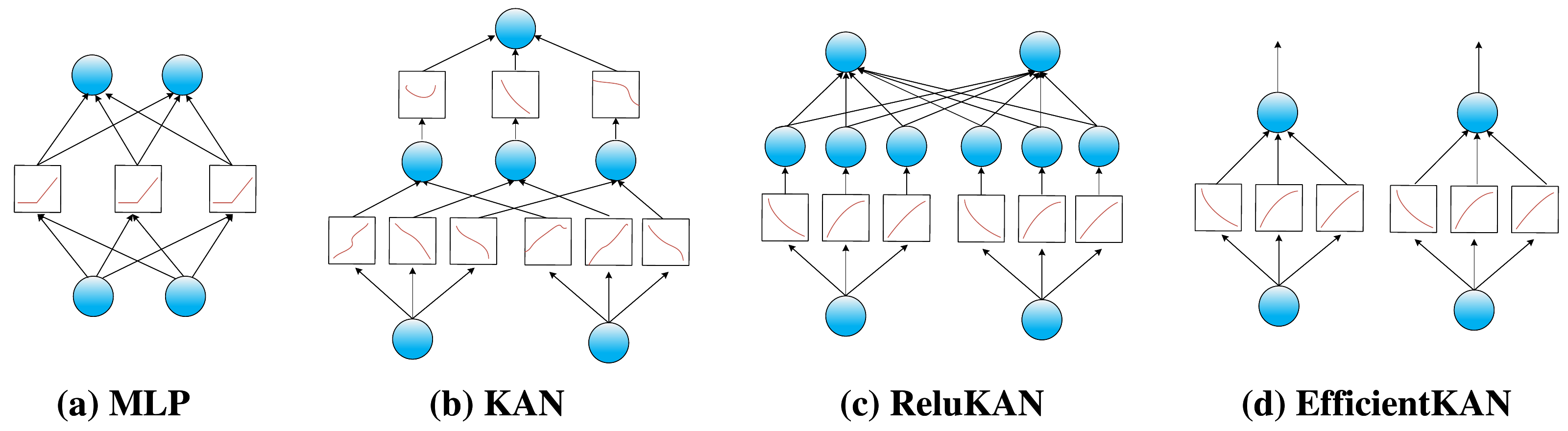}
  \caption{Structural Comparison Among Different Neural Networks: KAN revolutionizes the computational approach of multilayer perceptrons (MLPs) by placing learnable parameters within the activation operations, offering stronger nonlinear fitting capabilities but with computational challenges. ReluKAN simplifies the computational process but introduces a large number of parameters. EfficientKAN reduces both the parameter and computational of the original KAN.}
  \label{fig1}
\end{figure}

In recent years, U-Net \cite{UNet} and its variants \cite{UNet2plus,UNet3plus,AttentionUNet,nnUNet} have performed exceptionally well in medical image segmentation, driving continuous advancements in this field. For example, UNet++ \cite{UNet2plus} incorporates nested encoder-decoder sub-networks within the overall encoder-decoder network structure, redesigning the skip connections in the UNet architecture. UNet3+ \cite{UNet3plus} utilizes full-scale skip connections to directly combine high-level and low-level semantics from feature maps of different scales. It also employs deep supervisions to learn hierarchical representations from multi-scale aggregated feature maps. Methods like 3D U-Net and V-Net enhance performance by introducing three-dimensional convolutions to process 3D medical images \cite{3DU-Net,V-net}. However, these networks lack the ability to model long-range dependencies between features, indicating room for further improvement in the models. With the successful application of Transformers in computer vision, many Transformer-based networks have been introduced into medical image segmentation \cite{Transunet,Uctransnet,Swin-unet,FAT-Net,Transfuse,Medicaltransformer,H2Former}.  These methods model global dependencies within images, overcoming the limitations of traditional convolutional networks in handling long-range dependencies.

Despite the significant progress made by convolutional networks and Transformers in medical image segmentation, they still have inherent limitations. Specifically, convolutional operations primarily capture spatial dependencies between local pixels, making it challenging to effectively model complex nonlinear patterns across channels, which are often crucial for diagnosis in medical images \cite{FCPNet,PrivacyNet}. Transformers typically require large amounts of data for training, while medical image data is usually scarce. Additionally, Transformers have relatively weak capabilities in extracting local detail information, which is also essential in medical image segmentation \cite{he2023transformers,parvaiz2023vision}.

Recently, Kolmogorov–Arnold Networks (KANs) \cite{liu2024kan}, as an emerging network structure, have introduced learnable nonlinear activation functions, providing superior accuracy and interpretability, showing great potential in replacing traditional multilayer perceptrons (MLPs). KANs stack learnable nonlinear activation functions, enabling neural networks to learn complex functional mappings more efficiently, thereby improving model performance and interpretability.

To better balance the modeling capabilities of global and local information in medical image segmentation, we propose a new network architecture called TransUKAN. It combines the strengths of Convolutional Neural Networks (CNNs), U-Net, Transformers, and KANs. By introducing improved KAN into Transformers, TransUKAN enhances the modeling capability of local details while capturing global information. Additionally, the design of improved KAN significantly enhances the modeling of nonlinear relationships with only a small number of additional parameters, alleviating the training burden.

Our contributions can be summarized as follows:

1. We successfully complement the advantages of the KAN with Transformer and U-Net, using the local nonlinear modeling capability of the KAN to improve the Transformer structure. To the best of our knowledge, this is the first work to apply the KAN to medical image segmentation, providing a strong reference for subsequent research and application of KAN models in this field.

2. To address the issues of high memory usage and a large number of parameters in the KAN when processing images, we propose EfficientKAN. By sparsifying the matrices during the activation integration stage of the KAN, we simplify the computation process, making it efficiently applicable to medical image processing tasks.

3. We conducted extensive experimental validation of TransUKAN on multiple medical image segmentation tasks. Experimental results show that TransUKAN can achieve performance comparable to state-of-the-art methods with significantly reduced parameters, demonstrating its effectiveness and superiority in medical image segmentation tasks.

\begin{figure*}[t]
  \centering
    \includegraphics[width=\linewidth]{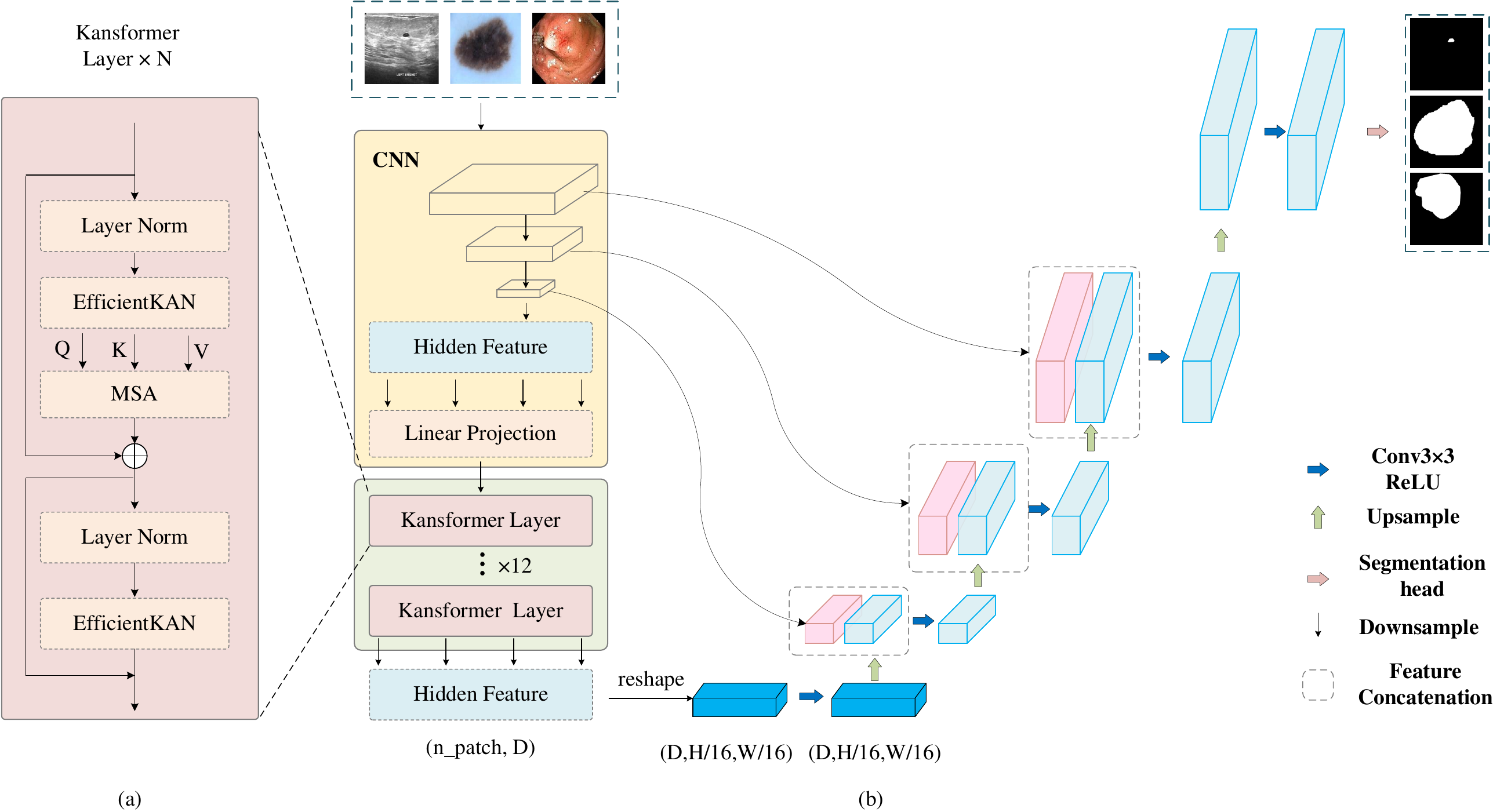}
  \caption{Overall structure of the TransUKAN. It consists of an encoder, decoder, and skip connections. An effective combination of KAN and Transformer was implemented in the encoder.}
  \label{fig2}
\end{figure*}

\section{Related Work}

\subsection{Medical Image Segmentation}

The advent of deep learning technologies, particularly CNNs, has significantly advanced the field of medical image segmentation. U-Net \cite{UNet} is one of the most classic CNN-based network architectures for medical image segmentation. The success of U-Net has inspired the development of numerous improved models. For instance, UNet++ incorporates nested encoder-decoder sub-networks and redesigned skip connections to improve feature fusion and gradient flow. UNet3+ leverages full-scale skip connections to combine multi-scale features and employs deep supervisions to learn hierarchical representations. 3D UNet and V-Net extend the traditional UNet architecture to three dimensions, making them suitable for volumetric image segmentation. Attention UNet and R2UNet introduce attention mechanisms to focus on relevant features, enhancing segmentation accuracy. Residual UNet and ResUNet++ incorporate residual connections to address the vanishing gradient problem, facilitating the training of deeper networks. Dense UNet and U-Net with Dense Blocks apply dense connectivity to improve feature reuse and network efficiency. Furthermore, UNetGAN combines UNet with generative adversarial networks (GANs) to improve segmentation quality through adversarial training. 

Furthermore, to address the lack of long-range dependency modeling capabilities in CNNs, researchers introduced Transformers into the field of computer vision, proposing the Vision Transformer (ViT). Subsequently, networks such as TransUNet, TransFuse, and UCTransNet emerged, combining Transformers with CNNs and fully exploring the extended capabilities of Transformers. Although Transformers can capture global information, they have a large number of parameters, high computational complexity, and are difficult to train. Additionally, Transformers are less effective at extracting local information, which can impact segmentation accuracy.

\subsection{Kolmogorov-Arnold Networks (KAN)}

Kolmogorov–Arnold Networks (KAN) is a neural network architecture based on the Kolmogorov–Arnold super approximation theorem. This theorem states that any multivariate continuous function can be represented as a finite superposition of univariate continuous functions. This theory provides a solid mathematical foundation for the representation and computation of complex functions. In KAN, by constructing a multi-layer network structure and utilizing the combination and superposition of univariate functions, complex nonlinear relationships can be efficiently approximated and represented. The core advantage of KAN lies in its ability to model nonlinear features efficiently with fewer parameters, making it highly effective in handling high-dimensional data and complex tasks. Specifically, KAN captures nonlinear patterns in input data through a series of linear transformations and nonlinear activation functions. Compared to traditional CNNs, KAN has advantages in terms of the number of parameters and computational complexity, making it highly applicable in scenarios requiring efficient models. 

In computer vision, ConvKANs \cite{bodner2024convolutional} adapt KANs into a convolutional architecture by integrating nonlinear activation functions from KANs into the convolutional layer. This integration effectively reduces parameter count while maintaining high accuracy levels. Graph-based applications also benefit from KAN \cite{kiamari2024gkan,zhang2024graphkan,bresson2024kagnns}, which replaces traditional MLPS in graph neural networks with KAN layers. By using learnable spline-based functions instead of fixed activation functions, this substitution enhances the model's ability to capture complex relationships in graph-structured data.

KANs offer significant advantages in accuracy and interpretability, positioning them as a promising alternative to traditional neural network models. However, the application of KAN in image processing has not been fully explored, especially in challenging areas such as medical image processing, where issues of high computation and memory usage have not been adequately addressed.

In our research, KAN has been improved and introduced into the task of medical image segmentation. By integrating it with the Transformer, we fully leverage its strengths in modeling nonlinear relationships. The improvements and introduction of KAN not only reduce the number of model parameters and lower computational complexity but also enhance segmentation accuracy. This combined approach effectively addresses the shortcomings of Transformers in local information extraction and significantly improves overall model performance and training efficiency.

\section{METHODOLOGY}

TransUKAN integrates the strengths of UNet, Transformer, and KAN, effectively combining them. We will first introduce the overall framework of the model in Section 3.1. Then, in Section 3.2, we will describe the integration of KAN and Transformer. Finally, in Section 3.3, we will elaborate on the detailed structure of the proposed EfficientKAN.

\subsection{Overall Structure}

As shown in Fig. 1, the overall structure of TransUKAN combines the advantages of CNN, Transformer, and KAN. First, the input image is processed by a CNN to extract features, generating feature maps. These feature maps are reshaped and linearly projected into a high-dimensional feature space. Then, these embedded features serve as the input to the Kansformer encoder, processed through multiple Kansformer layers to capture contextual information. Each Kansformer layer includes Layer Normalization (LN), Multi-Head Self-Attention mechanism (MSA), and EfficientKAN, ensuring the integration of global and local information in the features.

To restore the spatial resolution of the image, the encoded features are progressively upsampled through a cascaded upsampler. It consists of multiple upsampling blocks, each containing a 2x upsampling operation, a 3×3 convolutional layer, and a ReLU activation function. During the upsampling process, the encoded features are fused with high-resolution feature maps from the CNN encoding path through skip connections, enhancing the recovery of low-level spatial information for precise image segmentation. Finally, the fused feature maps are further upsampled to the full resolution of the original image to generate the final segmentation mask.

\subsection{Kansformer}

Since its inception, KAN has been optimized compared to the MLP. Its powerful nonlinear representation capability and training stability enable it to better capture the complex features of data. Therefore, an intuitive improvement approach is to directly replace MLP with KAN. Additionally, to maintain the continuity of nonlinear representation, we also replaced the QKV mapping matrices with a single-layer KAN. This allows the powerful nonlinear representation capability of KAN to be utilized in the self-attention mechanism, ensuring that the model maintains efficient nonlinear representation at each stage of feature extraction and processing, thereby improving overall performance.

The overall structure of Kansformer is shown in Figure 1(a), primarily composed of LN, EfficientKAN layers, and MSA. The output of the l-th layer can be expressed as follows:

\begin{equation}
\mathrm{z}_{l}^{\prime}=\operatorname{MSA}\left(\text{EfficientKAN}\left(LN\left(z_{l-1}\right)\right)\right)+z_{l-1}
\end{equation}
\begin{equation}
\mathrm{z}_{1}=\text{EfficientKAN}\left(LN\left(z^{\prime}\right)\right)+z^{\prime}
\end{equation}

Here, \( \text{LN}(\cdot) \) denotes the Layer Normalization operator, and \( z_l \) represents the feature map encoded by the Kansformer at layer \( l \).

By gradually replacing the MLP and QKV mapping matrices, the model benefits from KAN's parameter compression and nonlinear representation at each step. After replacing the MLP, the reduction in parameter count and increase in computational efficiency provides a solid foundation for further replacing the QKV mapping matrices. When the QKV mapping matrices are also replaced with KAN, the overall computational complexity is further reduced, and parameter efficiency is further improved. This synergistic optimization ensures that the model achieves optimal performance at each stage.

\subsection{EfficientKAN}

The core idea of KAN can be expressed by Eq. (3), which states that a high-dimensional function can be represented as a composition of a finite number of one-dimensional functions:

\begin{equation}
f(x) = \sum_{i=1}^{2n+1} \Phi_i \left( \sum_{j=1}^{n} \phi_{i,j}(x_j) \right)
\end{equation}

Specifically, assuming the input vector \( x \) has a length of \( n \), the computation of the output \( y \) can be expressed as follows:

\begin{equation}
y = \begin{pmatrix}
\Phi_{1} \\
\Phi_{2} \\
\vdots \\
\Phi_{2n+1}
\end{pmatrix}
\begin{pmatrix}
\varphi_{1,1} & \varphi_{1,2} & \cdots & \varphi_{1,2n+1} \\
\varphi_{2,1} & \varphi_{2,2} & \cdots & \varphi_{2,2n+1} \\
\vdots & \vdots & \ddots & \vdots \\
\varphi_{n,1} & \varphi_{n,2} & \cdots & \varphi_{n,2n+1}
\end{pmatrix} x
\end{equation}

Here, \( \phi_{i,j} \) is referred to as the inner function, and \( \Phi_i \) is referred to as the outer function. Specifically, the inner and outer functions can be expressed in the form of linear combinations and B-spline functions as follows:

\begin{equation}
\varphi(x) = w_b \frac{x}{(1 + e^{-x})} + w_s \sum c_i B_i(x)
\end{equation}

Here, \( B_i(x) \) is a B-spline function, \( w_b \) and \( w_s \) are weight parameters, and \( c \) is a control coefficient for shaping the B-spline. B-spline functions are a set of bell-shaped functions used to represent any univariate function on a finite domain. These functions have the same shape but different positions.

Due to the computational complexity of B-spline functions, KAN is limited in utilizing the parallel processing capabilities of GPUs. This complexity results in significant limitations in processing speed and scalability, especially in fine-grained classification tasks such as medical image segmentation. The ReLU-KAN architecture simplifies the basis function by adopting ReLU and dot product operations, optimizing the computational process as follows:

\begin{equation}
R_i(x) = \left[ \text{ReLU}(e_i - x) \times \text{ReLU}(x - s_i) \right]^2 \times \frac{16}{(e_i - s_i)^4}
\end{equation}

Here, \(e_i\) and \(s_i\) represent the upper and lower bounds of the basis function, respectively, controlling the range and position of the basis function through these two parameters. The factor \(\frac{16}{(e_i - s_i)^4}\) in the basis function plays a role in normalization and scaling. Its main purpose is to ensure that the activation value remains within a reasonable range, avoiding overflow or underflow. Subsequently, the computational process of KAN is optimized through convolution operations. This simplified basis function has higher computational efficiency, making it more suitable for GPU processing.

Through our experiments and observations, we found that although using convolution can make the KAN more favorable for GPU computation during the activation value integration, this method integrates not only the activation values of different neurons but also the activation values of individual neurons, treating all activation values as a whole. This approach can result in significant information redundancy and introduce a large number of parameters when handling high-dimensional data, leading to substantial computational resource usage during backpropagation. Specifically, assuming the input \( X \) has dimensions \( (B, C_{in}) \), where \( B \) is the batch size and \( C_{in} \) is the number of neurons, the network activated based on the ReLUKAN principle will generate \( X_1 \) with dimensions \( (B, (G+K), C_{in}) \), where \( G \) and \( K \) are hyperparameters used by KAN to generate the number of grids and also represent the height and width of the feature map and convolution kernel. The relationship between the activation values of each neuron calculated through convolution operations can be expressed as follows:

\begin{equation}
X' = \text{reshape}(X, (B, 1, G + K, C_{in}))
\end{equation}
\begin{equation}
X'' = \text{conv}(X', W)
\end{equation}

Here, the size of the convolution kernel is \( (G+K, C_{in}) \), and the dimension of the integrated feature map \( X'' \) is \( (B, C_{out}) \). This method introduces significant computational and parameter redundancy when handling high-dimensional data because the convolution kernel needs to integrate all neurons at each position:

\begin{equation}
\begin{aligned}
&\left[\begin{array}{ccc}
A_{11} & A_{12} & A_{13} \\
A_{12} & A_{22} & A_{23} \\
A_{13} & A_{32} & A_{33}
\end{array}\right]
\times
\left[\begin{array}{ccc}
B & C & D
\end{array}\right] \\
&\rightarrow
\left[\begin{array}{ccc}
A_{11} & \ldots & A_{33}
\end{array}\right]
\times
\left[\begin{array}{ccc}
B_{11} & C_{11} & D_{11} \\
\vdots & \vdots & \vdots \\
B_{33} & C_{33} & D_{33}
\end{array}\right] \\
&=
\left[\begin{array}{ccc}
X_{1} & X_{2} & X_{3}
\end{array}\right]
\end{aligned}
\end{equation}

Here, \( A_{ij} \) represents the activation values of different neurons in the input \( X \). The same row represents different activation values of the same neuron, and the same column represents different neurons. \( B \), \( C \), and \( D \) are convolution kernels that integrate the activation values of \( X \). ReluKAN replaces the original KAN activation value integration operation with convolution operations using kernels of the same size as the input feature map. Therefore, when the input feature map dimension increases, it leads to a significant increase in the number of parameters and computational load. To reduce computational and parameter redundancy, we limit the integration of activation values to within a single neuron and implicitly learn the relationships between different neurons only during backpropagation, as follows:

\begin{equation}
\begin{aligned}
&\left[\begin{array}{ccc}
A_{11} & \cdots & A_{1j} \\
\vdots & \ddots & \vdots \\
A_{i1} & \cdots & A_{jj}
\end{array}\right] \times
\left[\begin{array}{ccc}
a & \cdots & 0 \\
\vdots & \ddots & \vdots \\
k & \cdots & 0
\end{array}\right] \\
&= \left[\begin{array}{ccc}
X_{11} & \cdots & 0 \\
\vdots & \ddots & \vdots \\
X_{i1} & \cdots & 0
\end{array}\right] \rightarrow
\left[\begin{array}{ccc}
X_{1} & \cdots & X_{i}
\end{array}\right]
\end{aligned}
\end{equation}

By using this method, the activation value processing matrix is maximally sparsified. To further reduce parameters and computational load, the activation value processing matrix is further simplified such that \( a = k = 1/j \). In practical implementation, this can be replaced by an average pooling operation. Additionally, to further enhance the nonlinear fitting capability of KAN, a square calculation is introduced after the integration of activation values.

\section{EXPERIMENTS}

In this section, we test the effectiveness of our model using both external and internal datasets. Section 4.1 introduces the external and internal datasets. Section 4.2 details the experimental setup and procedures. Sections 4.3 and 4.4 present comparative experiments and ablation experiments, respectively, to demonstrate the effectiveness and superiority of TransUKAN.

\subsection{Datasets}

\textbf{ISIC} \cite{codella2018skin}: This dataset contains 2594 skin lesion images captured from real patients using a dermatoscope equipped with a digital camera. Each image has been annotated by a professional physician to mark the area of the skin lesion, and all data have been reviewed and managed by professional dermatologists with knowledge of dermatoscopy. 

\textbf{Kvasir} \cite{jha2020kvasir}: Kvasir-SEG is an open-source dataset manually annotated and verified by an experienced gastroenterologist. The dataset contains 1000 images of polyps and their corresponding masks, with image resolutions ranging from 332 $\times$ 487 to 1920 $\times$ 1072. The aim of creating this dataset is to promote the development and progress of polyp detection tasks.

\textbf{BUSI} \cite{BUSI}: The BUSI dataset collects ultrasound images of breasts from women aged between 25 and 75 years old. The dataset includes 780 images from 600 female patients, which are divided into three categories: normal, benign, and malignant. Among them, there are 133 normal cases, 437 benign tumors, and 210 malignant tumors. The average image size is 500 $\times$ 500 pixels.

\textbf{NKUT} \cite{zhou2024nkut}: NKUT is a specialized dataset designed for the segmentation of pediatric mandibular wisdom teeth (MWT) from Cone Beam Computed Tomography (CBCT) images. This dataset comprises 133 CBCT volumes, representing over 53,000 slices, with patient ages ranging from 7 to 22 years, and an average age of 13.2 years. The dataset includes detailed pixel-level annotations created by pediatric dentistry experts, covering bilateral MWT germs, second molars (SM), and partial alveolar bones (AB).

\subsection{Implementation Details}

The images in the dataset were uniformly preprocessed and adjusted to 256x256 pixels to meet the model's input requirements. The dataset was divided into training, validation, and test sets in an 8:1:1 ratio. Data augmentation techniques, including random cropping, rotation, and flipping, were widely applied during training to increase data diversity. Model training was conducted on a single NVIDIA A6000 GPU.

During training, we used the Adam optimizer with an initial learning rate set to 1e-4 and employed weight decay to prevent overfitting. The total training epochs were set to 200, with the first 10 epochs as a warm-up phase, using linear learning rate growth, followed by cosine annealing learning rate decay. Each training batch size was set to 8. For binary classification, the loss function is a weighted sum of cross-entropy loss and dice loss, while for multi-class classification, only cross-entropy loss is used. During the validation, we use DICE, IOU, and accuracy as evaluation metrics, and we also record model parameters as well as inference times to comprehensively evaluate the model's performance.

\subsection{Comparisons with Other SOTA models}

We conducted a comprehensive evaluation of our method against several state-of-the-art models across multiple datasets. The results are presented in Table \ref{table2}, it can be seen that our proposed method has significant competitiveness in medical image segmentation. Our model outperformed its counterparts across all six medical image segmentation tasks. These experimental results provide strong evidence that our method has effectively improved the capabilities of SAM in medical image segmentation. Furthermore, our method achieves the objective of employing a single model for segmenting multiple medical images, while consistently delivering excellent performance.

\begin{table*}[h]
    \caption{Quantitative comparison between SOTA methods and TransUKAN on Medical Image Datasets. All the metrics are based on the Acc performance.} 
    \centering 
    \begin{tabular}{l|@{\hspace{1em}}c|@{\hspace{1em}}c@{\hspace{1em}}c@{\hspace{1em}}c|@{\hspace{1em}}c@{\hspace{1em}}c@{\hspace{1em}}c}    
        \toprule 
        \multirow{2}{*}{Methods} & 
        \multirow{2}{*}{\#P(M)}&
        \multirow{2}{*}{BUSI} &
        \multirow{2}{*}{ISIC} & 
        \multirow{2}{*}{Kvasir} & 
        \multicolumn{3}{c}{NKUT} \\ 
        & &  &  &  & MWT & SAM & AB \\
        \midrule 
    UNet\cite{UNet}                           &17.27   &68.22 &89.98 &83.74 &57.01 &59.54 &28.03 \\
    Att-Unet\cite{AttentionUNet}              &34.88  &67.14 &89.86 &84.35 &64.07 &72.86 &52.96 \\
    TransUNet\cite{Transunet}                 &105.32  &72.76 &\textbf{91.58} &86.30 &89.67 &\textbf{90.13} &\textbf{80.94} \\
    TransDeeplab\cite{azad2022transdeeplab}   &17.49  &59.76 &89.06 &74.30 &85.75 &79.27 &75.59\\
    HiFormer\cite{heidari2023hiformer}        &23.25 &68.80 &91.13 &85.27 &62.44 &66.6 &44.78 \\
    UCTransNet\cite{Uctransnet}               &66.49 &71.49 &91.04 &86.02 &73.29 &70.67 &75.70 \\
    TransFuse\cite{zhang2021transfuse}        &26.28 &71.19 &90.55 &80.01 &69.06 &73.31 &50.56 \\
    AutoSAM\cite{hu2023efficiently}           &90.82 &70.04 &90.64 &82.58 &64.92 &69.18 &52.41 \\
    U-KAN\cite{li2024u}                       &25.36 &69.35 &90.47 &84.69 &42.24 &35.00 &25.26 \\
    \textbf{TransUKAN (Ours)}                 &20.85 &\textbf{75.46}      &91.17 & \textbf{87.75} & \textbf{90.29} &89.09 &77.96 \\
    \bottomrule         
\end{tabular}
\label{table1}
\end{table*}

\subsection{Ablation Studies}

In this section, we assess the impact of the proposed components on segmentation performance across external and internal datasets. All these models are based on TransUNet. The experimental results are shown in Table \ref{table3}.  The baseline model, comprising 105.3 M, demonstrates moderate performance on the external dataset with a DICE score of 91.8\%, 88.51\% and 75.58\%, respectively. 

\begin{table*}[h]
    \caption{TransUKAN ablation Studies on medical image datasets} 
    \centering 
        \begin{tabular}{l|@{\hspace{1em}}c|@{\hspace{1em}}c|@{\hspace{1em}}c@{\hspace{1em}}c@{\hspace{1em}}c|@{\hspace{1em}}c@{\hspace{1em}}c@{\hspace{1em}}c}    
        \toprule 
        \multirow{2}{*}{Methods} & 
        \multirow{2}{*}{\#P(M)}&
        \multirow{2}{*}{VM} &
        \multirow{2}{*}{BUSI} &
        \multirow{2}{*}{ISIC} & 
        \multirow{2}{*}{Kvasir} & 
        \multicolumn{3}{c}{NKUT} \\ 
        & & &  &  &  & MWT & SAM & AB \\
        \midrule 
        Vanilla  &105.3 &6 &72.76 &\textbf{91.58} &86.30 &89.67 &\textbf{90.13} &\textbf{80.94} \\
        +KAN     &21.23 &24 &73.78  &88.04  &73.78  &88.18  &86.35  &75.79 \\
        +ReLUKAN &233.2 &6 &73.15  &88.40  &74.19  &88.55  &86.62  &75.18 \\
        TransUKAN     &20.85 &6 &\textbf{75.46}      &91.17 & \textbf{87.75} & \textbf{90.29} &89.09 &77.96\\
        \bottomrule 
    \end{tabular}
\label{table3}
\end{table*}

After replacing all the fully connected operations in the Transformer with KAN, the number of parameters was reduced to 21.23 M, approximately five times lower. The results on external datasets remain highly competitive, demonstrating the potential of the KAN in medical image segmentation. However, after improving KAN to ReLUKAN, the introduction of numerous convolution operations led to an increase in model parameters to 233.2 M. Despite the increase in parameters, the model's performance did not significantly improve and even showed a downward trend, demonstrating that the ReLUKAN approach is not suitable for medical image processing tasks. Finally, using EfficientKAN to build the TransUKAN reduced the number of parameters to 20.85 M. This significantly lowered memory usage while accelerating training and improved the model's testing accuracy on external datasets, demonstrating the effectiveness of EfficientKAN.

Additionally, on the internal CBCT validation set, TransUKAN still achieved excellent performance, demonstrating its generalization and robustness.

\section{Conclusion}

In this study, we proposes TransUKAN, a medical image segmentation model based on EfficientKAN. By replacing the multi-layer perceptrons (MLP) and the QKV mapping matrices in the multi-head self-attention mechanism (MSA) of traditional Transformer models with Kolmogorov-Arnold Networks (KAN), the model's nonlinear representation capability, computational efficiency, and parameter efficiency are significantly improved. This study achieves a breakthrough in optimizing computational and parameter redundancy issues by using average pooling operations to integrate only the activation values of the current neuron, avoiding unnecessary computational burden while retaining key feature information. Experimental results demonstrate that the improved TransUKAN model significantly outperforms state-of-the-art models in terms of performance, while also significantly reducing computational complexity and the number of parameters, enhancing computational efficiency and model stability, thus validating its potential in practical applications.

Future research will further optimize the EfficientKAN structure, explore more efficient nonlinear basis function designs, and apply the model to multi-classification tasks and larger-scale datasets to validate its generality and robustness. Additionally, exploring the model's performance in real-time applications, optimizing inference speed and resource usage, will promote its application in practical medical scenarios, aiming to provide more advanced technical support for the field of medical image analysis and to advance the development of automated diagnostic technology.

\subsubsection{Acknowledgements} This work is partially supported by the National Natural Science Foundation (62272248), the China Scholarship Council (202306200119).

 \bibliographystyle{splncs04}
 \bibliography{mybib}
\end{document}